\documentclass{elsart}

 \usepackage{graphicx}

\usepackage{amssymb}

\newcommand{\jpsi}{J/$\psi$}
\newcommand{\psip}{$\psi^\prime$}
\newcommand{\jpsiDY}{J/$\psi$\,/\,DY}

\newcommand{\ezdc}{$E_{ZDC}$}

\begin{document}

\begin{frontmatter}

\title{\jpsi\ suppression in In-In collisions at 158 GeV/nucleon}
\vspace*{-0.2cm}

\author{Roberta Arnaldi$^{10}$} for the NA60 Collaboration:
\author{R.~Arnaldi$^{10}$, 
R.~Averbeck$^{9}$, 
K.~Banicz$^{2,4}$, 
J.~Castor$^{3}$},
\author{
B.~Chaurand$^{7}$,
C.~Cical\`o$^{1}$
A.~Colla$^{10}$, 
P.~Cortese$^{10}$,}
\author{
S.~Damjanovic$^{2}$,
A.~David$^{2,5}$,
A.~De~Falco$^{1}$, 
A.~Devaux$^{3}$,} 
\author{A. Drees$^{9}$,
L.~Ducroux$^{6}$,
H.~En'yo$^{8}$, 
A.~Ferretti$^{10}$, 
M.~Floris$^{1}$,}  
\author{A.~F\"{o}rster$^{2}$,
P.~Force$^{3}$,
N.~Guettet$^{3}$,
A.~Guichard$^{6}$, 
H.~Gulkanian$^{11}$, }
\author{J.~Heuser$^{8}$,
M.~Keil$^{2,5}$,
L.~Kluberg$^{7}$, 
C.~Louren\c{c}o$^{2}$, 
J.~Lozano$^{5}$, }
\author{F.~Manso$^{3}$, 
A.~Masoni$^{1}$,
P.~Martins$^{2,5}$, 
A.~Neves$^{5}$, 
H.~Ohnishi$^{8}$, } 
\author{C.~Oppedisano$^{10}$,
P.~Parracho$^{2}$, 
P.~Pillot$^{6}$, 
G.~Puddu$^{1}$, }
\author{E.~Radermacher$^{2}$,
P.~Ramalhete$^{2}$, 
P.~Rosinsky$^{2}$, 
E.~Scomparin$^{10}$, }
\author{J.~Seixas$^{5}$,
S.~Serci$^{1}$, 
R.~Shahoyan$^{2,5}$, 
P.~Sonderegger$^{5}$,}
\author{H.J.~Specht$^{4}$,
R.~Tieulent$^{6}$, 
G.~Usai$^{1}$, 
R.~Veenhof$^{5,2}$, 
H.K.~W\"ohri$^{1}$} 

\address{
$^{1}$Univ.\ di Cagliari and INFN, Cagliari, Italy,
$^{2}$CERN, Geneva, Switzerland,
$^{3}$LPC, Univ.\ Blaise Pascal and CNRS-IN2P3, Clermont-Ferrand, France,
$^{4}$Univ.\ Heidelberg, Heidelberg, Germany,
$^{5}$IST-CFTP, Lisbon, Portugal,
$^{6}$IPN-Lyon, Univ.\ Claude Bernard Lyon-I and CNRS-IN2P3, Lyon, France,
$^{7}$LLR, Ecole Polytechnique and CNRS-IN2P3, Palaiseau, France,
$^{8}$RIKEN, Wako, Saitama, Japan,
$^{9}$SUNY Stony Brook, New York, USA,
$^{10}$Univ.\ di Torino and INFN, Italy,
$^{11}$YerPhI, Yerevan, Armenia
}
\vspace*{-0.2cm}

\begin{abstract}
The NA60 experiment has studied \jpsi\ production in Indium-Indium collisions at 158 A$\cdot$GeV.
In this paper we present an updated set of results obtained with the complete set of available statistics and  
an improved alignment of the vertex tracker. 
The centrality dependence of the \jpsi\ production, obtained with an analysis
technique based only on the \jpsi\ sample, indicates that a suppression beyond that induced by nuclear absorption is present in In-In
collisions, setting in at $\sim$80 participant nucleons. A first study of the systematic errors related
with this measurement is discussed. 
We also present preliminary results on the \jpsi\ azimuthal distributions.
\end{abstract}

\end{frontmatter}

\vspace{-0.7cm}
\section{Introduction}
\label{intro}
\vspace{-0.5cm}
The study of dilepton production in heavy-ion collisions at the CERN SPS provides some of the most
interesting observables explored so far in the search for the quark-gluon plasma. In particular \jpsi\ suppression is one of 
the most important signatures for the formation of a deconfined phase \cite{Satz86}. 
The NA38 and NA50 experiments have extensively studied \jpsi\ production in various colliding systems, including 
p-A, S-U and Pb-Pb \cite{Abreu05}. 
Proton-nucleus data provide an important reference, which describes the 
expected absorption for \jpsi\ crossing cold nuclear matter. By comparing the centrality dependence of the \jpsi\ yield 
observed in A--A collisions to this reference, one can look for anomalous \jpsi\ suppression. 
In particular, NA50 has observed, in Pb-Pb collisions, that below a certain centrality threshold
the \jpsi\ production is well described invoking nuclear absorption as the only suppression mechanism; on the contrary, above that
 threshold, an anomalous suppression sets in, and the \jpsi\ yield becomes considerably 
lower than expected from the nuclear absorption.
The observations done by NA38/NA50 surely give a clear indication for an anomalous \jpsi\ suppression, 
however they should be complemented by other sets of accurate measurements obtained from different collision systems.
The NA60 experiment at CERN SPS, with an innovative experimental apparatus improving the dilepton
detection technique, studies \jpsi\ production in
In-In collisions, in order to investigate if the suppression is already present in systems lighter than
Pb-Pb. The comparison between different colliding systems should also allow to understand which is the
physics variable at the origin of the \jpsi\ suppression as a function of centrality.

\vspace{-0.7cm}
\section{The experimental apparatus}
\label{setup}
\vspace{-0.5cm}
The NA60 experimental apparatus is based on the muon spectrometer inherited from the NA50 experiment,
complemented by a new target region, based on a Beam Tracker and a Vertex Tracker placed in a 2.5 T dipole
magnet. 
The target system is made of 7 In targets, 1.5 mm thick each, placed in vacuum, corresponding to a
total interaction probability of $\sim$20\%.
A Zero Degree Calorimeter (ZDC) provides an estimate of the centrality of the collisions, by 
measuring the energy (\ezdc) released by the projectile nucleons which have not taken part to the interactions.
A more detailed description of the apparatus can be found in \cite{Usai,Keil}.
The reconstruction of the muon tracks in the muon spectrometer is affected by multiple scattering and
energy loss fluctuations due to the crossing of the muons through the hadron absorber. To overcome this
limitation, NA60 measures muons before their entrance into the hadron absorber, by means of the vertex
tracker. Muon tracks measured in the muon spectrometer are matched with tracks reconstructed by the vertex tracker, 
resulting in an improvement of the dimuon mass
resolution and in an accurate determination of the muons' origin. Details on the matching technique can be
found in \cite{RubenQM}. 

\vspace{-0.7cm}
\section{\jpsi\ results from the Indium-Indium run}
\label{results}
\vspace{-0.5cm}
During the 5 weeks long In-In run in 2003, NA60 has collected $\sim$230 million dimuon
triggers, running with beam intensities around 5$\times10^7$ per 5-second spill. 
Two muon spectrometer settings have been used during the data taking, corresponding to two values of the magnet current: 4000A and 6500A.
Events collected with the lower current have a better acceptance for the low mass dimuons, while the
higher current setting improves the mass resolution of the \jpsi. 
With respect to the previously published results \cite{ArnaldiQM05}, the analysis presented in this paper
includes both data sets, i.e. the whole statistics collected by the NA60 experiment.
Furthermore, a new reconstruction of the data has been performed. 
It is based on a better alignment of the vertex tracker, resulting in few ${\mu}$m accuracy, 
and on an improved quality of the tracking algorithm in the vertex spectrometer. Details on these issues can be found in \cite{ADavidHP}.

The event selection criteria have been tuned in order to obtain a clean sample of \jpsi\ events produced in primary In-In collisions.
Tracks are reconstructed in the vertex tracker and we require to have at least one vertex satisfying various quality cuts, including a
minimum number of tracks ($\ge 4$) attached to it.
The longitudinal coordinate of that vertex, which is reconstructed with an accuracy of $\sim$200 ${\mu}$m, must lie inside one of the seven
targets. 
The matching of the muon tracks makes it possible to determine the longitudinal coordinate of the 
dimuon production point with a precision of few hundreds ${\mu}$m.
Asking the dimuon vertex to coincide with the most upstream interaction vertex in the target region ensures 
that the \jpsi\ is produced in a primary In-In collision. 
A further set of selections is done in order to minimize the event pile-up in our apparatus. 
It is based on the good timing properties (a few ns) and on the excellent granularity of the Beam Tracker, that allows to reject events with 
several ions arriving close in time. This selection is particularly important for obtaining an unbiased \ezdc\ distribution. 
In fact, since the ZDC is placed on the beam axis and records a signal for each incoming ion, it is particularly sensitive to this kind of background.
Finally, in order to reject events coming from the edges of the spectrometer acceptance we only retain dimuons corresponding to the phase space 
window $0 < y_{cms} <1$ and $-0.5 < \cos\theta_{CS} < 0.5$, where  $\theta_{CS}$ is the polar angle of the muons in the Collins-Soper reference 
frame. 

Two complementary analysis have been performed in order to study the centrality dependence of the \jpsi\ 
production in Indium-Indium collisions. They differ in the way used for normalizing the \jpsi\ distributions. 
The first technique, the so-called ``standard'' analysis, is based on the normalization of the \jpsi\ to the Drell-Yan (DY) 
yield. In this approach, already used by the NA38/NA50 experiments, Drell-Yan is used as a reference process. 
In fact, being independent from final state effects, its production cross section scales with the number of collisions and represents, therefore, an
ideal normalization for a hard process, like \jpsi\ production.
In the ``standard'' analysis, the ratio between the \jpsi\ and the DY yields is extracted
by fitting the opposite sign dimuon mass spectrum to a superposition of different contributions. 
Above 2 GeV/c$^2$, the dimuon invariant mass region contains the \jpsi\ and the \psip\ resonances, an underlying continuum composed
of Drell-Yan events and semi-muonic decays of  D and $\overline{\rm D}$ mesons, and a combinatorial
background from $\pi$ and K decays. 
The expected mass shapes of the signals are evaluated through a Monte Carlo simulation based on Pythia with GRV94LO parton
distribution functions. These simulations also provide the values of the \jpsi\ and DY acceptances ($A_{{\rm
J}/\psi}$=12.4\,\%, $A_{DY(2.9-4.5)}$=13.0\% for the data collected with the current of 6500A in the muon 
spectrometer and slightly larger for the 4000A).
The combinatorial background has been estimated from the measured sample of like-sign pairs.
Results obtained from the two sets of data, corresponding to the two configurations of the muon
spectrometer, are found to be statistically compatible. Therefore in the following we show results based on their average.
The study of the ratio between \jpsi\ and DY cross-sections has the advantage of being free from systematic
errors connected to experimental inefficiencies and to the integrated luminosity, since both \jpsi\ and DY events were collected
under the same conditions. However, as it was mentioned,
the statistical error is large, due to the limited number of high mass DY. Therefore, in order to increase the statistics as much as possible, the
matching of the muon candidates in the vertex spectrometer has not been required. 
This allows to increase the statistics by $\sim$40\%, for a total of $\sim$320 DY events with mass above 4.2 GeV/c$^2$ and $\sim$45000 \jpsi. 
In this case a further cut based on the extrapolation of the muon spectrometer tracks ensures that the dimuon comes from the
target region. In spite of the statistics gain, the number of high mass Drell-Yan events is still limited; as a consequence the \jpsiDY\ ratio 
cannot be studied in more than three centrality bins. 
\begin{figure}[hbt]
\centering
\includegraphics[width=0.45\textwidth]{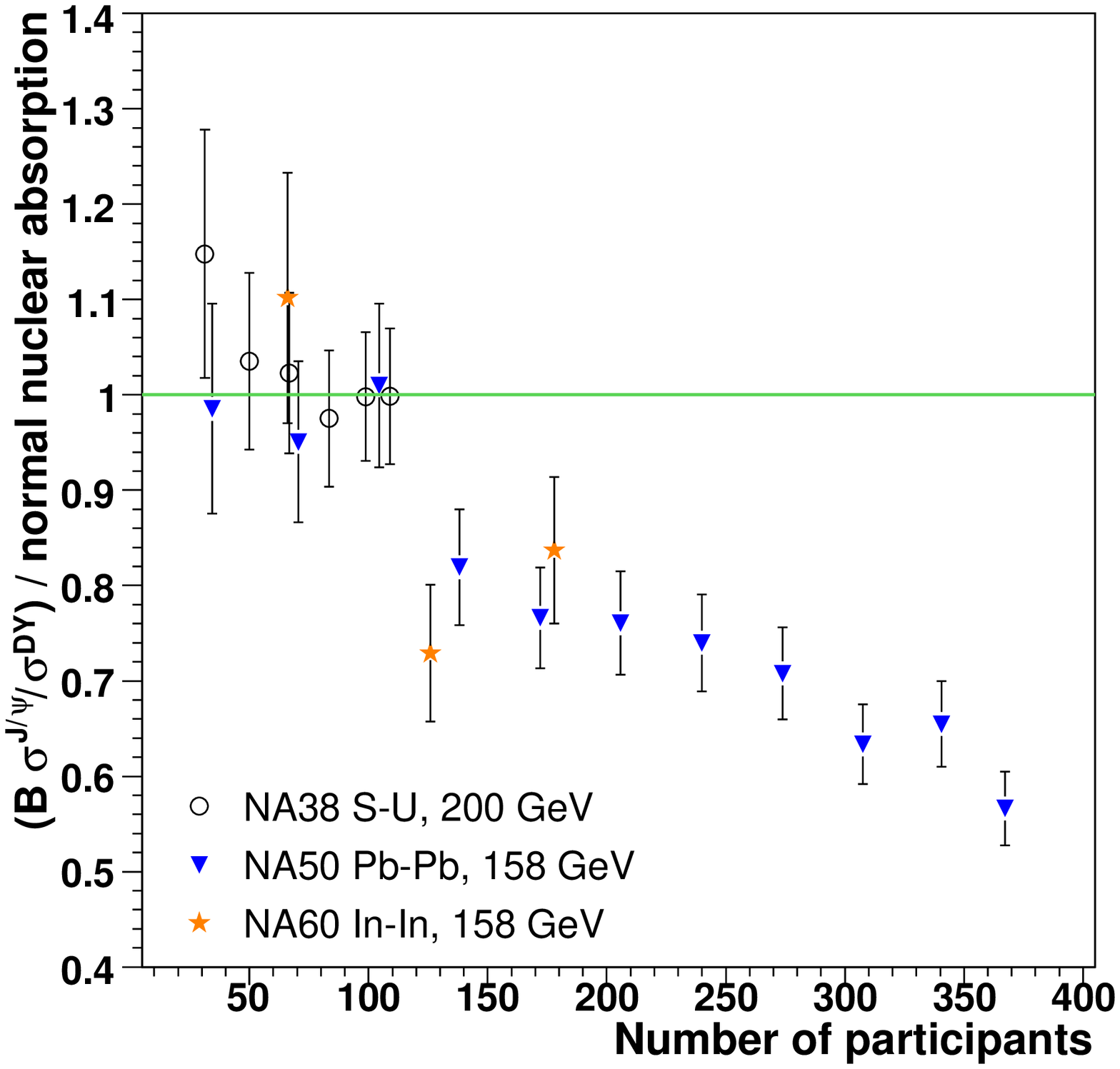}
\includegraphics[width=0.45\textwidth]{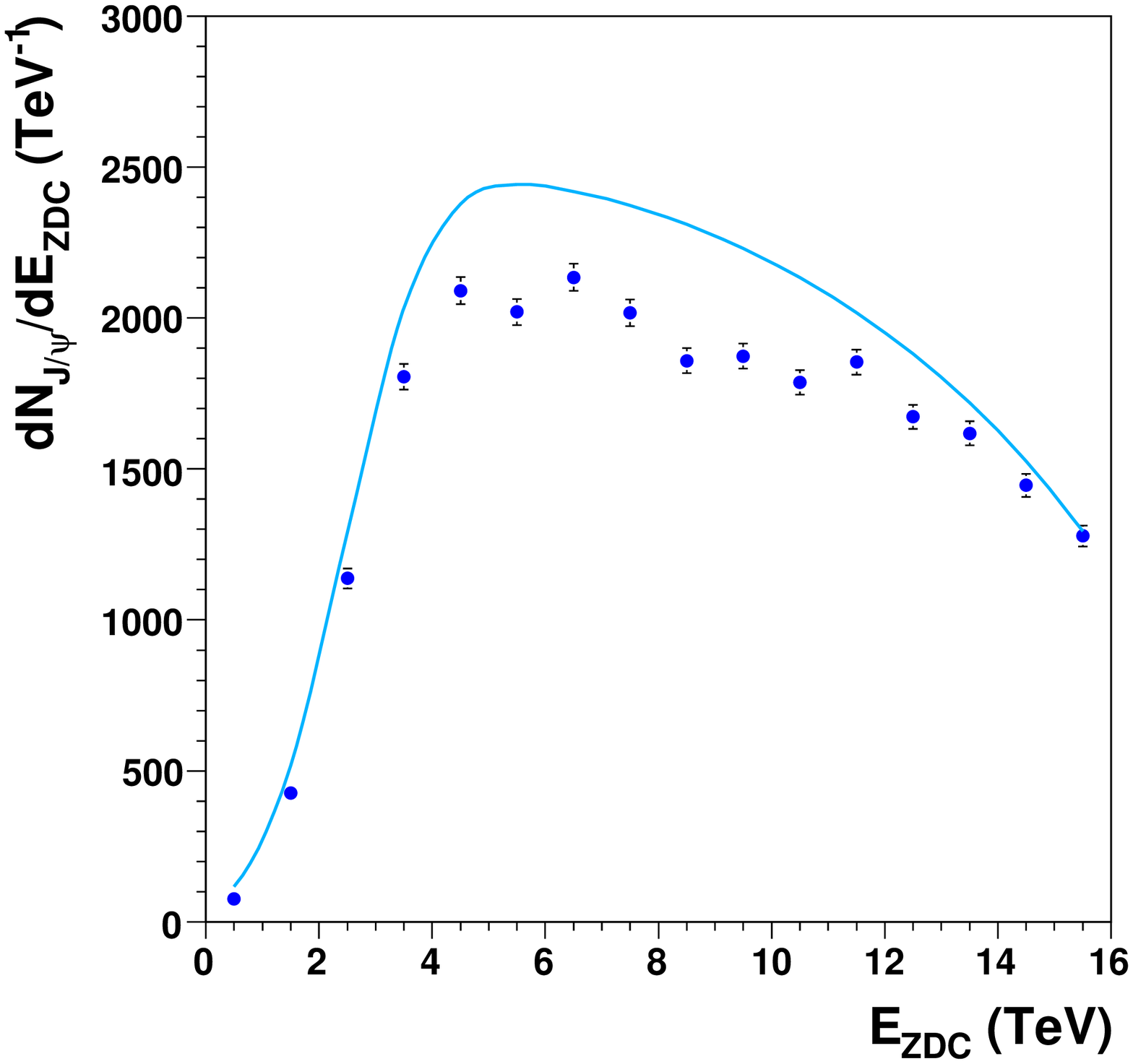}
\caption{Left: \jpsiDY\ standard analysis divided by the normal nuclear absorption curve vs. N$_{part}$. NA50
and NA38 results are also shown. Right: measured \jpsi\ compared to the normal nuclear absorption curve vs. \ezdc. The error on the relative normalization between the absorption
curve and the measured points is not shown.}
\label{fig:JPsi}
\end{figure}

The \jpsiDY\ values have to be compared to expectations from a pure nuclear absorption scenario (the so-called ``nuclear absorption curve'').
This reference distribution is obtained using the Glauber model, assuming that \jpsi\ production is a hard process and adopting the value
$\sigma^{abs}_{J/\psi}$=4.18 $\pm$ 0.35 mb, as measured by NA50 \cite{Abreu05}, for the nuclear
absorption cross section of the c$\overline{\rm c}$ state.
In Fig. \ref{fig:JPsi} (left) the measured \jpsiDY\ ratios have been divided out by such a curve and plotted as a function of the 
number of participants (N$_{part}$). 
This centrality estimator has been obtained from \ezdc\ using the Glauber model and taking into account the smearing induced by the detector 
resolution.
We see that a suppression of the \jpsi\ beyond nuclear absorption is present in 
In-In collisions. However, the small number of bins prevents an accurate study of the \jpsi\ centrality
dependence as well as a meaningful comparison with the results obtained by NA38/NA50.

The second analysis technique overcomes the problem of the low Drell-Yan statistics by directly studying the measured \jpsi\ centrality
distribution as a function of \ezdc. Following this approach, the \jpsi\ sample is not normalized to a measured quantity. 
Rather, the observed \jpsi\ centrality distribution is directly compared to the theoretical distribution expected in case nuclear 
absorption is the only suppression mechanism.
Since the number of \jpsi\ is not normalized to other data collected under the same
conditions, one has to check that no centrality-dependent inefficiencies are introduced by the selection criteria and, more
generally,
that systematic errors are reduced as much as possible. 
For this reason in this analysis we require the matching of the muon tracks in order to identify and discard events where the \jpsi\ is not produced
in primary In-In collisions. We have checked, by means of a Monte Carlo simulation, that the matching efficiency as a function of centrality for 
\jpsi\ events increases by less than 2\% from central to peripheral events, inducing a negligible bias on the 
centrality distribution of the measured \jpsi. 
In addition, the matching improves significantly the \jpsi\ mass resolution, from 105 to 70 MeV, and reduces the combinatorial background from 3 
to 1\% in the corresponding mass region.
We have also checked that the vertexing efficiency for events where a \jpsi\ has been produced, decreases by less than 1\% from central to 
peripheral collisions. 
After event selection, the \jpsi\ distribution as a function of centrality has been obtained, in 1 TeV \ezdc\ bins, by means of a simple fitting 
procedure that allows to subtract the small amount of Drell-Yan and combinatorial background under the resonance peak.
The \jpsi\ centrality distribution is then compared to expectations from pure nuclear absorption. The relative normalization between data and 
the reference curve is not fixed a priori; therefore we assume the ratio between data and the nuclear absorption curve,
integrated over centrality, to be the same as in the ``standard'' \jpsiDY\ analysis, i.e. 0.87 $\pm$ 0.05.
\begin{figure}[hbt]
\centering
\includegraphics[width=0.45\textwidth]{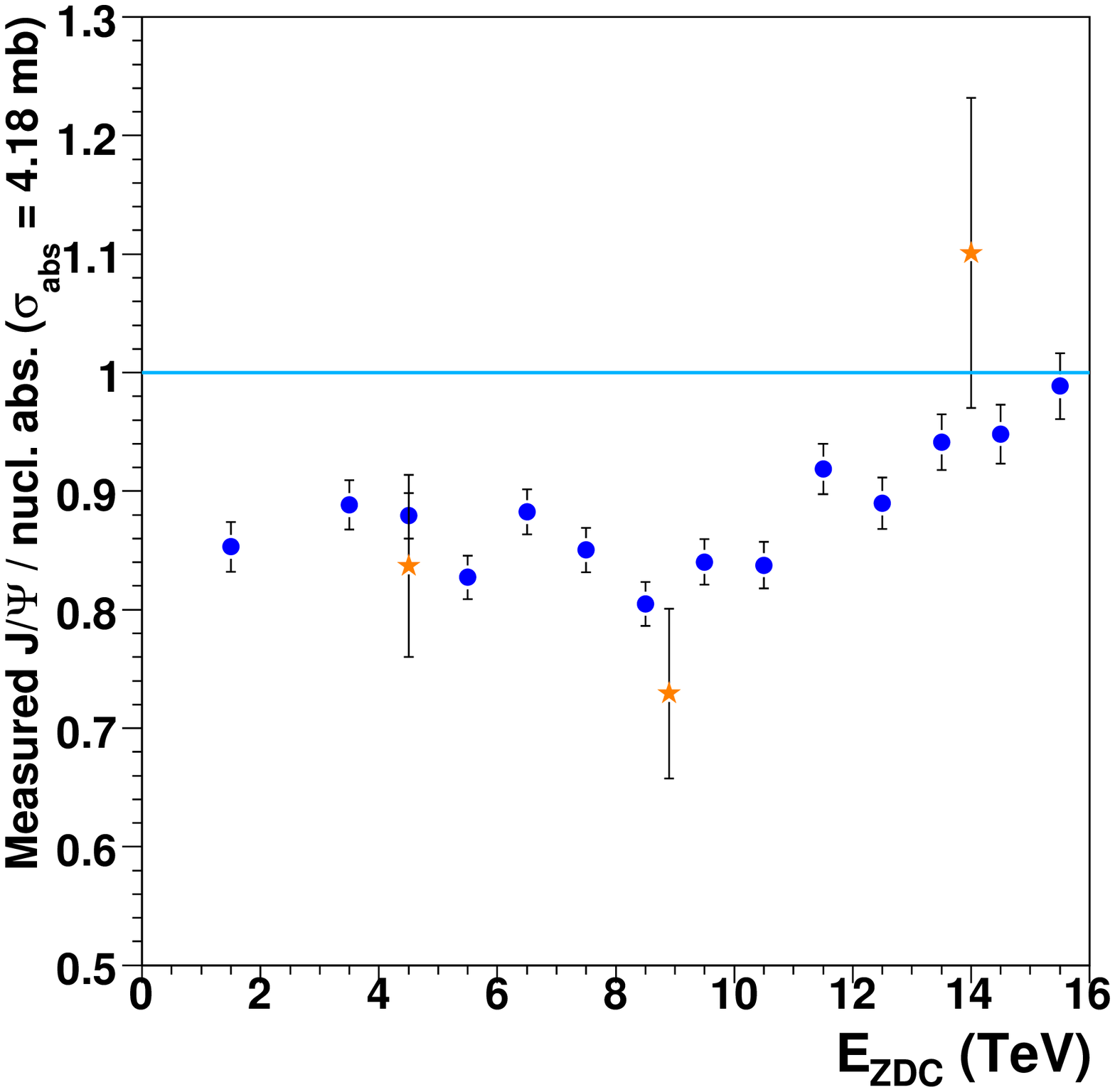}
\includegraphics[width=0.45\textwidth]{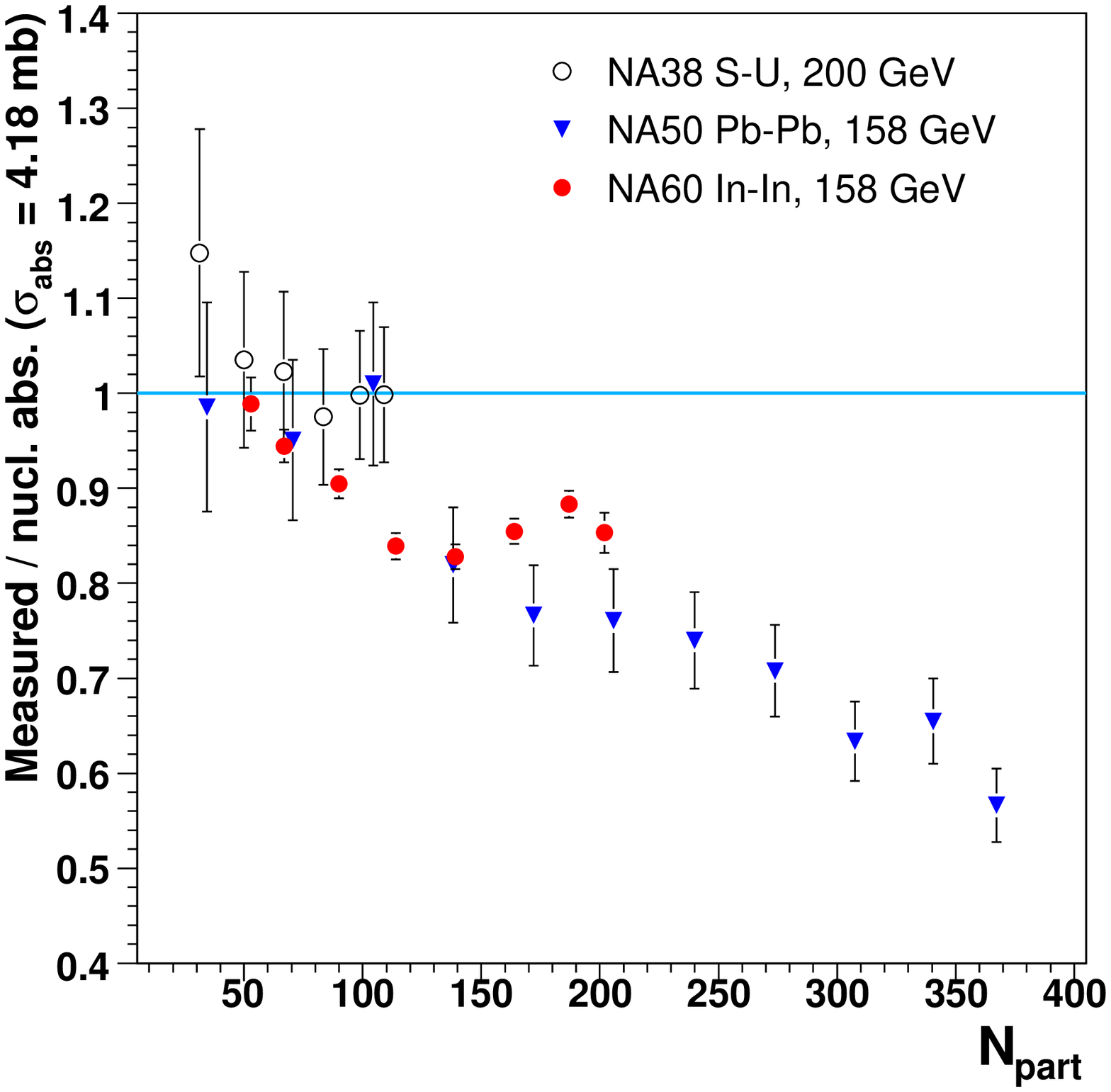}
\caption{Left: ratio between measured \jpsi\ and the absorption curve vs. \ezdc\ (stars: standard
analysis). The three most central bins of Fig. \ref{fig:JPsi} (right) have been grouped together, 
to reduce statistical fluctuations. Right: ratio between measured \jpsi\ and the absorption curve 
as a function of N$_{part}$ for S-U, Pb-Pb and In-In collisions.}
\label{fig:JPsi2}
\end{figure}

The result is shown in Fig. \ref{fig:JPsi} (right) as a function of \ezdc, while in Fig. \ref{fig:JPsi2} (left) we plot the ratio between the measured and the 
expected \jpsi\ yield. It must be noted that events where a heavy nuclear fragment produced in a primary In-In collision reinteracts in a downstream 
target have a biased \ezdc\ value. A Monte-Carlo correction (smaller than 2\%) has been introduced to correct for the apparent shift towards central 
events induced by this effect.
This analysis confirms the presence of an anomalous \jpsi\ suppression in In-In collisions, already seen in the ``standard'' approach.
The departure from the expected normal nuclear absorption occurs at $\sim$12 TeV and is followed by a rather flat behaviour 
for more central events. 
In Fig. \ref{fig:JPsi2} (right) the In-In results, rebinned in order to reduce statistical fluctuations, are plotted
together with the previous results of NA38/NA50 as a function of the number of participant nucleons. Although this comparison must be considered 
preliminary, due to the possible systematics connected with the use of different centrality estimators (transverse energy for NA38/NA50, \ezdc\ for
NA60), a fair agreement between the various systems can be observed.
As a side remark, it can be noticed that the In-In suppression pattern shown in Fig.~\ref{fig:JPsi2} (right) is compatible with the occurence of
a step-like suppression as a function of N$_{part}$, once the smearing due to the \ezdc\ resolution is taken into account.
Such a kind of study shows that the departure from the normal nuclear absorption behaviour occurs at N$_{part}$ $\sim$80.       
As can be seen in Fig.~\ref{fig:JPsi2} this approach provides results with negligible statistical errors ($<2$\%). 
Concerning systematic errors, we have verified that their most important source is connected with the determination of the nuclear absorption curve 
and with the determination of N$_{part}$ starting from \ezdc. 
The calculation of the absorption curve is based on the Glauber model. Therefore we have investigated the influence of the choice of the inputs on the
result. Different Indium nuclear density distributions \cite{Landolt} with respect to the default parameterization
\cite{DeVries} have been tested. This choice does not affect significantly the shape of the absorption curve, since the effect is smaller than 1\% for
\ezdc$>3$ TeV. For very central events the effect is larger and reaches $\sim$15\%. 
Concerning the centrality determination, the zero degree energy is well correlated with N$_{part}$.
However, the ZDC does not measure only spectator nucleons but also a small amount of energy released by the forward secondary particles, emitted in 
the calorimeter's acceptance ($\eta > 6.3$). 
The uncertainty on this contribution, important only for very central collisions, has been conservatively assumed to be of the order of 10\%.
We have calculated that such an uncertainty gives, for events with \ezdc\ $< 3$ TeV, a systematic error of the order of 9\% on the absorption curve.
For more peripheral events the effect is negligible.
Furthermore, the calculation of the absorption curve depends on ${\sigma}^{pp}_{J/\psi}$  and on ${\sigma}^{abs}_{J/\psi}$, both estimated in p-A 
collisions at 450 GeV by NA50. The experimental errors on such values, and in particular the factors needed to rescale ${\sigma}^{pp}_{J/\psi}$ 
from 450 to 158 GeV, induce a systematic error on the absorption curve of the order of 8\%, almost independent of centrality.
Finally, since we have normalized the data to the absorption curve using the results from the ``standard'' analysis, the 6\% error related to the
uncertainty in the normalization factor has also to be taken into account.
Combining the various sources, we end up with a $\sim$10\% systematic error, independent of centrality. On top of that, the most 
central bin (\ezdc\ $<3$ TeV) is affected by a further, sizeable systematic error relatively to the others. 
This means that the shape of the suppression pattern has been determined with good precision, while the absolute normalization of the points is more 
uncertain. It must be noted, anyway, that a significant fraction of the systematic errors originates from the uncertainty in the energy rescaling of 
the absorption curve. The p-A data collected by NA60 at 158 GeV should help to reduce significantly this uncertainty. 
Finally, one should be aware that the systematic errors related to the energy rescaling of the absorption curve also affect the NA38/NA50 points. 
Therefore they do not influence the relative positions of the patterns observed in S-U, In-In and Pb-Pb.

\begin{figure}[hbt]
\centering
\includegraphics[width=0.43\textwidth]{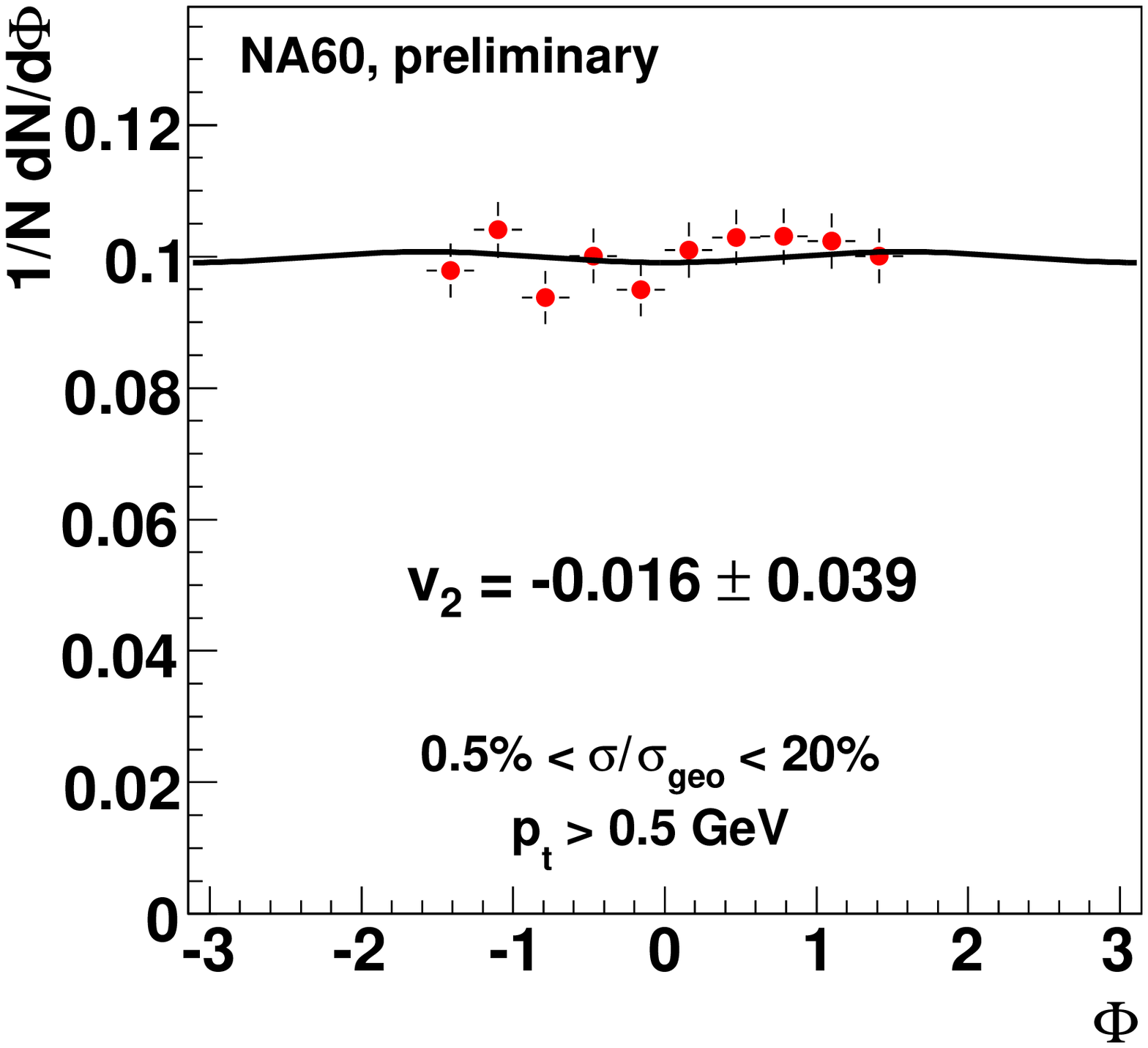}
\includegraphics[width=0.43\textwidth]{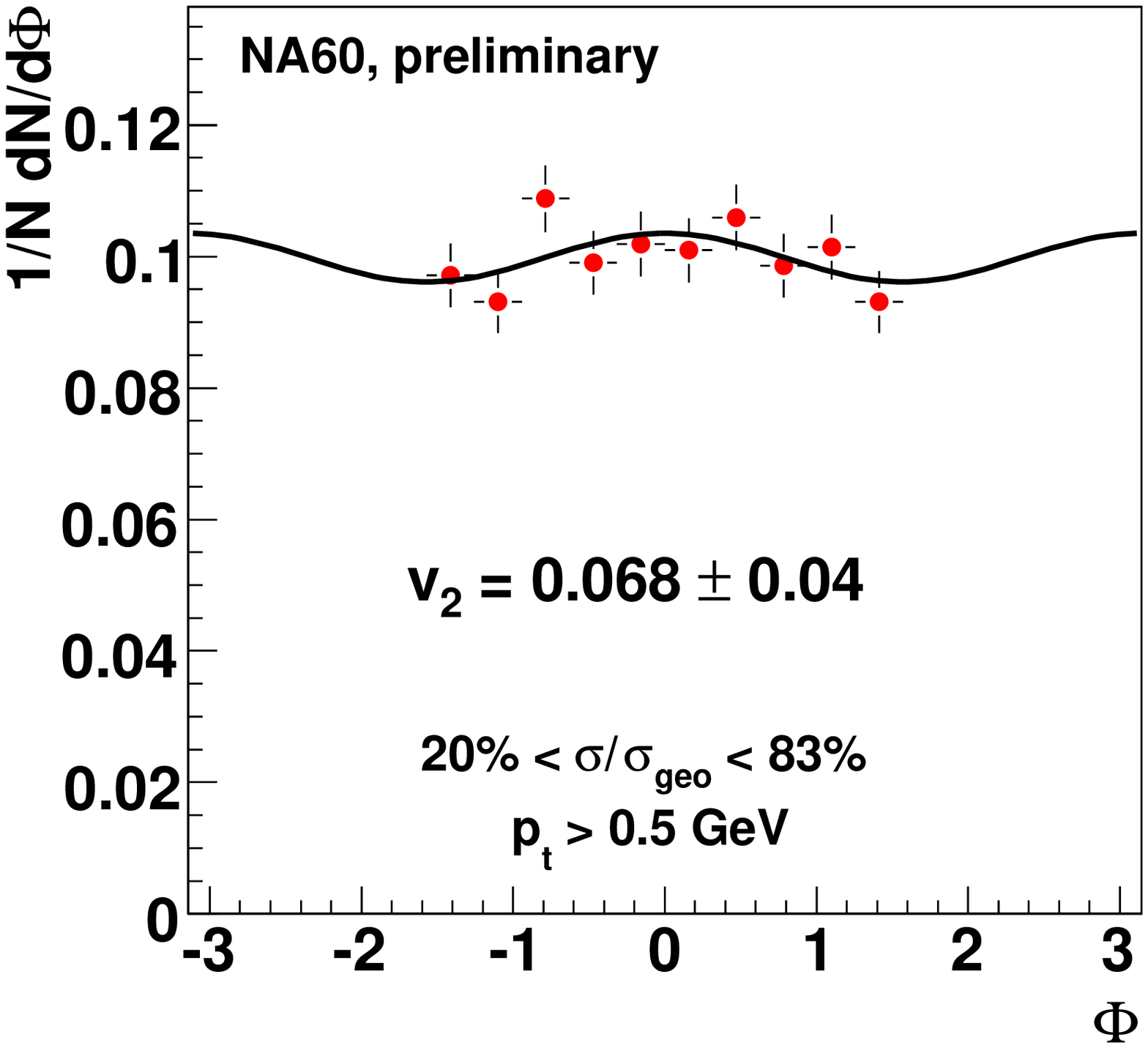}
\caption{\jpsi\ azimuthal pattern for (left) central and (right) peripheral collisions.}
\label{fig:jpsiflow}
\end{figure}

NA60 has also obtained preliminary azimuthal angle distributions of the \jpsi\ produced in In-In collisions. These distributions have been obtained 
with respect to the orientation of the reaction plane, defined by the beam direction $\vec{\rm z}$ and the impact parameter vector $\vec{\rm b}$.
This plane is experimentally not directly accessible, but it can be determined from the emission
angles of the charged particles measured in the vertex tracker, using the so-called event plane method.
The \jpsi\ anisotropic flow is then quantified evaluating the coefficients of a Fourier expansion of the azimuthal distribution. 
The azimuthal angle distribution for \jpsi\ is shown in Fig. \ref{fig:jpsiflow} (left) for 
central events ($ 0.5 < \sigma/\sigma_{geo} <20$\%) and in Fig. \ref{fig:jpsiflow} (right) for peripheral
events  ($ 20 < \sigma/\sigma_{geo} <83$\%). 
The elliptic flow coefficient ${\rm v_2}$ is obtained after correcting for the event
plane resolution. 
The distributions are based on $\sim$12000 \jpsi, which is about $\sim$50\% of the full
statistics available for this kind of analysis. 
The more peripheral data seems to indicate a non-isotropic emission pattern but
the limited statistics does not allow, for the moment, strong conclusions.
A detailed review of this analysis is reported in \cite{Forster}.  

\section{Conclusions}
\label{conclusions}
\vspace{-0.5cm}
The analysis of the data taken by NA60 in In-In collisions shows that the \jpsi\ is anomalously suppressed. 
An analysis method based on the comparison of the \jpsi\ centrality distribution with the expectation from a pure nuclear absorption scenario
shows that the anomalous suppression sets in at N$_{part}\sim$80, with a saturation for more central events. The statistical errors are negligible 
(of the order of 2\%). We have estimated a 10\% systematic error on this analysis, independent of centrality and therefore not affecting the shape 
of the \jpsi\ suppression pattern. Most of the systematic errors come from the uncertainty in our knowledge of the \jpsi\ nuclear absorption, which 
has been accurately measured at 450 GeV, but not at  158 GeV, the energy of the heavy-ion data taking. The situation is expected to improve as soon 
as results from the analysis of the p-A data collected by NA60 at 158 GeV will be available.

\end{document}